\documentclass[a4paper,oneside,final,notitlepage,onecolumn]{article}
\usepackage{amsfonts}
\usepackage{epsf}
\usepackage{t1enc}
\usepackage[latin2]{inputenc}

\DeclareFontFamily{OT1}{rsfs}{} \DeclareFontShape{OT1}{rsfs}{m}{n}{
<-7> rsfs5 <7-10> rsfs7 <10-> rsfs10}{}
\DeclareMathAlphabet{\mycal}{OT1}{rsfs}{m}{n}

\begin{document}

\newtheorem{theorem}{Theorem}[section]
\newtheorem{lemma}{Lemma}[section]
\newtheorem{proposition}{Proposition}[section]
\newtheorem{corollary}{Corollary}[section]
\newtheorem{conjecture}{Conjecture}[section]
\newtheorem{example}{Example}[section]
\newtheorem{definition}{Definition}[section]
\newtheorem{remark}{Remark}[section]
\newtheorem{exercise}{Exercise}[section]
\newtheorem{axiom}{Axiom}[section]
\newtheorem{condition}{Condition}[section]
\renewcommand{\theequation}{\thesection.\arabic{equation}} 

\author{\small 
István Rácz\thanks{%
~email: iracz@sunserv.kfki.hu}
\\ 
\small MTA KFKI, Részecske- és Magfizikai Kutatóintézet\\
\small  H-1121 Budapest, Konkoly Thege Miklós út 29-33.\\ 
\small Hungary\\ }

\date{\small \today}

\title{{\bf On the use of the Kodama vector field in \\ spherically
    symmetric dynamical problems}} 

\maketitle

\begin{abstract}
It is shown that by making use of the Kodama vector field, as a
preferred time evolution vector field, in spherically symmetric
dynamical systems unexpected simplifications arise. In particular, the
evolution equations relevant for the case of a massless scalar field
minimally coupled to gravity are investigated. The simplest form of
these equations in the 'canonical gauge' are known  to possess the
character of a mixed first order  elliptic-hyperbolic system. The
advantages related to the use of the Kodama vector field are
two-folded although they show up simultaneously. First, it is found
that the true degrees of freedom separate. Second, a subset of the
field equations possessing the form of a first order symmetric
hyperbolic system for these preferred degrees of freedom is singled
out. It is also demonstrated, in the appendix, that the above results
generalise straightforwardly to the case of a generic
self-interacting scalar field.
\end{abstract}

\parskip 5pt

\section{Introduction}
\setcounter{equation}{0}

Let us start with a four dimensional spherically symmetric dynamical
spacetime. The metric of such a spacetime can be given as 
\begin{equation}
{\rm d}s^2=g_{AB}{\rm d}x^A{\rm d}x^B+ R^2({\rm d}\theta^2+\rm
sin\theta^2 {\rm d}\phi^2), \label{le0} 
\end{equation}
where $g_{AB}$ is a non-singular Lorentzian two-metric, moreover,
$g_{AB}$ and $R$ are smooth functions of the coordinates $x^A$
(the capital Latin indices take the values $1$ and $2$) exclusively.  
The Kodama vector field \cite{ko} (see also \cite{hay}, and also
\cite{misao} for a generalisation to higher dimensional spacetimes) is
defined as 
\begin{equation}
K^a=\epsilon^{ab}{\partial_a}R, \label{kv0} 
\end{equation}
where $\epsilon^{ab}$ denotes the volume form associated with the
two-metric $g_{AB}$.
It can be checked by a straightforward calculation that 
\begin{equation}
\nabla^e K_e=0,
\end{equation}
moreover, by making use of the Einstein's equations it can also be
shown (see for details \cite{ko}) that 
\begin{equation}
G^{ef}\nabla_e K_f
\end{equation}
does vanish, where $G_{ab}$ denotes the Einstein tensor. This
later relation implies that the vector field 
\begin{equation}
J^a =G^{ab}K_b
\end{equation}
is also divergence free. This means that even though the spacetime might be
completely dynamical, i.e. there might be no timelike Killing vector
field in the underlying setting, $J^a$ is a locally
conserved energy flux vector field.

\parskip 5pt

\section{Dynamical spacetimes with a massless scalar field}
\setcounter{equation}{0}
 
Let us consider now the particular case of a massless scalar field
minimally coupled to gravity. The Einstein equations are given as
\begin{equation}
G_{ab}=8 \pi T_{ab}\label{Ein}
\end{equation}
with energy-momentum tensor\footnote{%
Notice that in the particular case of the massless
scalar field considered here 
$\nabla^a  \left(T_{ab} K^b\right)$
does also vanish, as it should do, whenever all the independent field
equations -- including (\ref{we}) --  are satisfied. } 
\begin{equation}
T_{ab}=\nabla_a \psi \nabla_b \psi -\frac12
g_{ab}\nabla^e \psi \nabla_e \psi . \label{Tab}
\end{equation}
The field equation relevant for the scalar field reads as 
\begin{equation}
\nabla^e \nabla_e \psi=0.\label{we}
\end{equation}
It follows from (\ref{Ein}) and (\ref{Tab}) that 
\begin{equation}
\nabla^a G_{ab}=8 \pi \nabla_b \psi (\nabla^e \nabla_e \psi)
\end{equation}
as it should be in consequence of the diffeomorphism invariance of the
underlying theory.

Hereafter, we shall use a regular spherical coordinate system
$(t,r,\theta,\phi)$ so that $r$ is chosen to be the area radial
coordinate, i.e. a two-sphere of radius $r$ has proper area 
\begin{equation}
\mathcal{ A}=4\pi r^2. \label{ac}
\end{equation} 
Then the canonical form  of the line element can be given as 
\begin{equation}
{\rm d}s^2=-A {\rm d}t^2+B {\rm d}r^2+ r^2({\rm d}\theta^2+\rm
sin\theta^2 {\rm d}\phi^2), \label{le} 
\end{equation}
where the metric functions $A$ and $B$ are assumed to be smooth
functions of the coordinates $t$ and $r$.
Notice that in these coordinates the Kodama vector field takes the
form 
\begin{equation}
K^a=\frac{1}{\sqrt{AB}}\left(\frac{\partial}{\partial
  t}\right)^a. \label{kv} 
\end{equation}
 
To recognise the intimate relation between the Einstein equations, in
particular which of them can be derived from the others it is informative
to consider the divergence  
\begin{equation}
\nabla^a E_{ab}=\nabla^a \left(G_{ab}-8 \pi T_{ab}\right)
\end{equation}
of the tensorial expression
$E_{ab}=G_{ab}-8 \pi T_{ab}$ vanishing of which implies that the
Einstein equations are satisfied. Using now the fact that $E_{ab}$ has
only four non-zero (algebraically independent) components one gets that
the only two non identically vanishing components -- the $t$ and $r$
components -- of $\nabla^a E_{ab}$ can be given as 
\begin{eqnarray}
&& \hskip-1cm-2\,A{B}^{2}r \left( {\frac {\partial}{\partial t}}{\it E}_{{tt}}
  \right)+2\,{B}^{2}r{\it E}_{{tt}}{\frac {\partial }{\partial t}}A
  \mbox{}+ABr{\it E}_{{tr}} \left( {\frac {\partial }{\partial r}}A
  \right) +2\,{A}^{2}Br \left( {\frac {\partial }{\partial r}}{\it
  E}_{{tr}} \right) \nonumber \\ &&
\hskip-1cm\phantom{-2\,A{B}^{2}r \left( {\frac {\partial}{\partial
  t}}{\it E}_{{tt}} 
  \right)} -ABr{\it E}_{{tt}} \left( {\frac {\partial
  }{\partial t}}B \right)   -{A}^{2}r{\it E}_{{tr}}{\frac {\partial
  }{\partial r}}B-{A}^{2}r{\it E}_{{rr}}{\frac {\partial }{\partial
  t}}B \mbox{}+4A^2B\,{\it E}_{{tr}}=0\,, \label{Et}
\end{eqnarray}
\begin{eqnarray}
&& \hskip-1.5cm2\,A{B}^{2}{r}^{3} \left( {\frac {\partial}{\partial t}}{\it
    E}_{tr} \right) -{B}^{2}{r}^{3}{\it E}_{{tt}}{\frac
    {\partial}{\partial r}}A
\mbox{}-{B}^{2}{r}^{3}{\it E}_{tr}{\frac {\partial}{\partial
    t}}A+AB{r}^{3}{\it E}_{tr} \left( {\frac {\partial}{\partial t}}B
    \right) -AB{r}^{3}{\it E}_{{rr}} \left( {\frac {\partial}{\partial
    r}}A \right) \nonumber \\ &&
    \hskip-1.5cm\phantom{2\,A{B}^{2}{r}^{3} \left( {\frac
    {\partial}{\partial t}}{\it 
    E}_{tr} \right)}-2\,{A}^{2}B{r}^{3} \left( {\frac {\partial}{\partial
    r}}{\it E}_{{rr}} \right) 
\mbox{}+2\,{A}^{2}{r}^{3}{\it E}_{{rr}}{\frac {\partial}{\partial
    r}}B-4\,{A}^{2}B{r}^{2}{\it E}_{{rr}}+4\,{A}^{2}{B}^{2}{\it
    E}_{{\theta\theta}}=0\,, \label{Er}
\end{eqnarray}
respectively.

The standard way of interpretation goes as follows. If one assumes
that the field equations $E_{rr}=0$ and 
$E_{\theta\theta}=0$ are satisfied one can get by the substitution of
these relations to (\ref{Et}) and (\ref{Er}) a first order linear
homogeneous system of PDEs relevant for  $E_{tt}$ and $E_{tr}$. This
system possesses the trivial solution for trivial data which implies
that the "constraints propagate", i.e. $E_{tt}=0$ and $E_{tr}=0$ are
satisfied everywhere if they hold on the initial data surface.

In the present case there is an other possible way of
interpretation of the above relations, as well. Assume now that the
field equations $E_{tt}=0$, $E_{tr}=0$ 
and $E_{rr}=0$ are satisfied. The substitution of these relations to
(\ref{Er}) implies that the fourth Einstein
equation follows from the others, i.e. $E_{\theta\theta}=0$ is
automatically satisfied.

Since the other relation (\ref{Et}) has not been used here there has
remained a further possible simplification between the Einstein
equations $E_{tt}=0$, $E_{tr}=0$ and $E_{rr}=0$. To determine it
assume now that only the field equations $E_{tt}=0$ and $E_{rr}=0$ are
satisfied everywhere. The substitution of these relations into
(\ref{Et}) yields then   
\begin{equation}
ABr{\it E}_{{tr}} \left( {\frac {\partial }{\partial r}}A \right)
+2{A}^{2}Br \left( {\frac {\partial }{\partial r}}{\it E}_{{tr}}
\right) -{A}^{2}r{\it E}_{{tr}}{\frac {\partial }{\partial r}}B 
\mbox{}+4{A}^{2}B\,{\it E}_{{tr}}=0,
\end{equation}
which is a linear PDE for $E_{tr}$ and can be considered as ODEs for
$E_{tr}$ on each of the time level surfaces. These equations possesses
the trivial solution, i.e. $E_{tr}=0$ is also satisfied everywhere,
whenever $E_{tr}=0$ is satisfied at the centre.

By combining the above two statements one can conclude, as it was done by
Choptuik \cite{cho} (see also \cite{gu}), that all of the
algebraically independent Einstein equations are satisfied whenever
$E_{tt}=0$ and $E_{rr}=0$ hold 
everywhere and, in addition, $E_{tr}=0$ is guaranteed to be satisfied
at the centre.

\section{The basic structure of the field equations}
\setcounter{equation}{0}

By making use of the auxiliary variables 
\begin{equation}
\Phi=\frac {\partial}{\partial r} \psi,\ \ \ \ \Pi=\sqrt{\frac
  BA}\frac {\partial }{\partial t}\psi \label{sub1}
\end{equation}
(\ref{we}) can be given in the form of the following first-order
system 
\begin{equation}
{\frac {\partial }{\partial t}}\,\Phi ={\displaystyle \frac {
\sqrt{{
\frac {A}{B}}}  \left(\,r\,({\frac {\partial }{\partial r}}\,\Pi ) +
\Pi (1-\,B) \right)}{\,r}}\label{1st}
\end{equation}
\begin{equation}
{\frac {\partial }{\partial t}}\,\Pi ={\displaystyle \frac {
\sqrt{{
\frac {A}{B}} }\,\left(\Phi (1+ \,B)  + \,r \,(
{\frac {\partial }{\partial r}}\,\Phi )\right)}{r}} \label{2nd}
\end{equation}
Moreover, the Einstein equations $E_{tt}=0$, $E_{tr}=0$ and $E_{rr}=0$
can be shown to take the form 
\begin{equation}
{\frac {\partial }{\partial r}}\,B= 
\frac {B(1-B)}{r} +4\,\pi \,r\,B\left( \Pi ^{2} +\,\Phi ^{2}\right)\label{Ett}
\end{equation}
\begin{equation}
{\frac {\partial }{\partial t}}\,B=
  8\,\pi  \,r\,\Phi \,\Pi \,\sqrt{{
 {A}{B}} }\label{Etr}
\end{equation}
\begin{equation}
{\frac {\partial }{\partial r}}\,A= -\frac{A (1- B)}{r} + 4\,\pi
\,r\,A\left(\,\Pi ^{2} +\,\Phi ^{2}\right). \label{Err} 
\end{equation}
Because all the variables $A,B, \Phi$ and $\Pi$ appears in the above
equations by adapting Choptuik's approach one could simply evolve the
scalar field according to the hyperbolic equations (\ref{1st}) and (\ref{2nd}) 
and solve the elliptic equations (\ref{Ett}) and (\ref{Err}) on each time level
surface. According to the argument concerning the dependence of the
Einstein equations on each other this method provides a completely
satisfactory solution for the full elliptic-hyperbolic system of
equations provided that 
(\ref{Etr}) is also ensured to be satisfied at the origin. This
later property can easily be guaranteed by the obvious choice $B=1$ at
the centre which also necessary to avoid the presence of a conical
singularity there. It is a remarkable property of the gauge choice
applied by Choptuik that all the equations (\ref{Ett}) - (\ref{Err})
concerning the gravitational degrees of freedom are of first order
ones, moreover, two of these equations, along with the assumption that
$B=1$ at the centre, determine the functions $A$ and $B$ completely.   

\section{The Kodama vector field as a time evolution vector field}
\setcounter{equation}{0}

The significance of the use of the Kodama vector field in relation
with the above discussed dynamical problem gets to be transparent upon
applying the Kodama vector field, as the time evolution vector field,
instead of $(\partial/\partial t)^a$ which was used, e.g., in the
above described approach of Choptuik. To see this replace all the time
derivatives with 
respect to the coordinate $t$ by derivatives with respect to a
variable $\tau$ in the above given field equations according to the
rule
\begin{equation}
\frac{\partial}{\partial t}=\sqrt{AB}\frac{\partial}{\partial
  \tau}\,. \label{r1}
\end{equation} 
We can also think of the variable $\tau$ as a natural parametrisation
of the Kodama vector field with respect to which it can be given as
$K^a=(\partial/\partial \tau)^a$. 

We also need to replace the $r$-derivatives in the above field
equations because of the simultaneous change of the coordinates
$(t,r)$ into $(\tau,\rho)$ given by the implicit relation  
\[
\tau=\tau(t,r)\ \ \ {\rm and} \ \ \ \rho=r.
\]
Then for an arbitrary function $F$ given originally as a $C^1$
function of $(t,r)$ we have the relations 
\begin{equation}
\frac{\partial}{\partial \rho}F=\frac{\partial}{\partial
  r}F+\frac{\partial}{\partial t}F \frac{\partial t}{\partial
  \rho}\,,\label{s1} 
\end{equation}
\begin{equation}
\frac{\partial}{\partial \tau}F=\frac{\partial}{\partial t}F
\frac{\partial t}{\partial \tau}, \label{s2}
\end{equation}  
where by (\ref{r1}) for the coefficient ${\partial t}/{\partial \tau}$
we have  
\begin{equation}
\frac{\partial t}{\partial \tau}=\frac{1}{\sqrt{AB}} \label{s3}
\end{equation}
but we do not know the other coefficient ${\partial t}/{\partial
  \rho}$. In fact, it is not this factor what we really need. Instead,
as it is implied by (\ref{s1}) and (\ref{s2}) 
\begin{equation}
\frac{\partial}{\partial r}F=\frac{\partial}{\partial
  \rho}F- \frac{\partial t}{\partial
  \rho} \left(\frac{\partial t}{\partial
  \tau}\right)^{-1}  \frac{\partial}{\partial \tau}F\,,\label{r2} 
\end{equation}
whence only the knowledge of the combined factor $C=\frac{\partial
  t}{\partial \rho} \left(\frac{\partial t}{\partial
  \tau}\right)^{-1}= \sqrt{AB}\frac{\partial
  t}{\partial \rho}  $ is needed to get the field equations relevant for
  the the introduced new variables. This, however, also means that the
  function $C$ will explicitly be present in the evolution equations for
  $B, \Phi$ and $\Pi$, therefore, we shall need an evolution equation
  for $C$, as well. Such an equation can be derived as follows. In
  virtue of  (\ref{s3}) we have that $\sqrt{AB}\frac{\partial
  t}{\partial \tau}=1$ holds. By deriving this relation  with respect
  to $\rho$ we get then that 
\begin{eqnarray}
&&\hskip-1cm \frac{\partial }{\partial
  \rho}\left(\sqrt{AB}\frac{\partial t}{\partial 
  \tau}\right)=\frac{\partial }{\partial
  \rho}\left(\sqrt{AB}\right)\frac{\partial t}{\partial
  \tau}+\sqrt{AB}\left(\frac{\partial
  t}{\partial \tau \partial \rho}\right)\nonumber \\
  &&\hskip-1cm\phantom{\frac{\partial 
  }{\partial \rho}\left(\sqrt{AB}\frac{\partial t}{\partial 
  \tau}\right)}=\frac{\partial t}{\partial
  \tau} 
  \left(\frac{\partial }{\partial \rho}\sqrt{AB}
  -\left[\frac{\partial t}{\partial 
  \rho}\left(\frac{\partial t}{\partial
  \tau}\right)^{-1}\right]\frac{\partial }{\partial \tau}\sqrt{AB}
  \right)+\frac{\partial }{\partial
  \tau}\left(\sqrt{AB}\frac{\partial t}{\partial  
  \rho}\right)=0\,,
\end{eqnarray}
which in virtue of (\ref{s3}), (\ref{r2}), (\ref{Ett}) and (\ref{Err})
yields the desired evolution equation for $C$ as 
\begin{equation}
\frac{\partial}{\partial \tau}C=-\frac{1}{\sqrt{AB}}
\frac{\partial}{\partial r}\sqrt{AB} = -\frac12\left[ 
\frac1A\frac{\partial A}{\partial r} + \frac1B\frac{\partial
  B}{\partial r}\right]= -4\,\pi
\,\rho\,\left(\,\Pi ^{2} +\,\Phi ^{2}\right)\,.\label{c} 
\end{equation} 
It can also be seen that the replacements (\ref{r1}) and (\ref{r2})
yields from equations (\ref{1st}), (\ref{2nd}) and (\ref{Etr})
the relations
\begin{equation}
{\frac {\partial }{\partial \tau }}\,\Phi ={\displaystyle 
\frac { \,B\,({\frac {\partial }{\partial \rho }}\,\Pi )\, - 
\,C\,({\frac {\partial }{\partial \rho }}\,\Phi )}{ B^{2} - C^{2}}
} + {\displaystyle \frac {\,B(B -1)\,\Pi  -  \,C (B+1)\,\Phi } {\,\rho\,
    ( B^{2} - C^{2})}}\,,\label{11st} 
\end{equation}
\begin{equation}
{\frac {\partial }{\partial \tau }}\,\Pi ={\displaystyle \frac {
 B\,({\frac {\partial }{\partial \rho }}\,\Phi ) - C\,({\frac {
\partial }{\partial \rho }}\,\Pi )}{ B^{2} - C^{2}}}  
+ {\displaystyle \frac {\,B(B +1)\,\Phi  -\,C (B-1)\,\Pi } {\,\rho( B^{2} -
    C^{2})}}\,, \label{22nd}
\end{equation}
\begin{equation}
{\frac {\partial }{\partial \tau}}\,B=
  8\,\pi  \,\rho\,\Phi \,\Pi. \label{eEtr}
\end{equation}
Notice that the metric variable $A$ does not appear in either of the
equations (\ref{c}), (\ref{11st}), (\ref{22nd}) and
(\ref{eEtr}). Moreover, they consist of a first order symmetric
hyperbolic system 
for which the initial value problem is known to be well-posed
\cite{ch}. What is even more important is that this system of 
evolution equations, along with 
\begin{equation}
\,{\frac {\partial }{\partial \rho }}\,{\rm ln}A - C\,{\frac {
\partial }{\partial \tau }}\,{\rm ln}A  - \frac{(\,B  -1)}{\rho}  - 4\,\pi
\,\rho\,\left(\,\Phi ^{2}+ \,\Pi ^{2}\right)=0\,, \label{A}
\end{equation}
derived from (\ref{Err}), consist of a complete system, i.e.  
\begin{equation}
\frac {\partial }{\partial \rho }\,B=4\,\pi\,\rho  \,\left(B\,(\Phi ^{2} + 
\Pi ^{2}) + 2\,C\,\Pi\,\,\Phi \right) - {\displaystyle \frac {B\,(B
 - 1)}{\rho}}, \label{const}
\end{equation}
derived from (\ref{Ett}) and $E_{\theta\theta}=0$ are automatically
satisfied whenever (\ref{c}), (\ref{11st}), (\ref{22nd}), (\ref{eEtr}) and
(\ref{A}) hold everywhere, moreover, (\ref{const})
is satisfied on the initial data surface.  Notice also that since the
equations (\ref{c}) -- (\ref{eEtr}) separates from (\ref{A}) the later
one can be solved independently upon the functions $B, C, \Phi$ and
$\Pi$ are determined. 

Thus we solve the Cauchy problem relevant for the chosen dynamical system
by specifying first initial data for the variables $B, C, \Phi$ and
$\Pi$ so that they satisfy the constraint (\ref{const}). We let then the
system to evolve under the control of the hyperbolic equations (\ref{c}),
(\ref{11st}), (\ref{22nd}) and (\ref{eEtr}). Finally, the metric
variable $A$ has to be determined by solving the linear PDE (\ref{A})
in terms of the already known functions
$B, C, \Phi$ and $\Pi$.

We would like to emphasise that the apparently singular
behaviour of equations (\ref{11st}), (\ref{22nd}), (\ref{A}) and
(\ref{const}) at the centre, $\rho=0$, is compensated by the fact that
whenever the functions $B$ and $\Phi$ are smooth they can be given
in a neighbourhood of the centre as  
\begin{equation}
B=1+\beta \rho^2\ \ 
{\rm and }\ \ \Phi=\phi \rho,
\end{equation}
with $\beta
$ and $\phi$ being smooth functions of $\tau$ and
$\rho$.

It is also important to know whether the expression
$B^2-C^2$ appearing in the denominators of (\ref{11st}) and
(\ref{22nd}) may vanish anywhere. In this respect we have only the
following simple observations. It seems to be reasonable to have
initial data $C\equiv 0$ on the initial data surface which choice
corresponds to the case where the $t=const$ and $\tau=const$ surfaces
coincide at our initial data surface. Notice, moreover, that by 
(\ref{c}) the value of $C$ cannot increase during the evolution so it
will be always non-positive. In
addition, we also have by (\ref{c}) and (\ref{eEtr}) that   
\begin{equation}
\frac{\partial}{\partial \tau}(B+C)=-4\pi\,\rho\,(\Pi-\Phi)^2
\end{equation}
\begin{equation}
\frac{\partial}{\partial \tau}(B-C)=4\pi\,\rho\,(\Pi+\Phi)^2.
\end{equation}
The second of these equations implies that $B-C$ will never
vanish since $B-C=1$ on the initial data surface. However,
the first of these equation does not exclude the vanishing of $B+C$.
It may happen that $B+C$ vanishes somewhere in the computation
domain. Note, however, that the vanishing of $B+C$ would correspond to 
a coordinate singularity of the metric given in the new coordinates
as 
\begin{equation}
{\rm d}s^2=-\frac1B {\rm d}\tau^2-2\,\frac CB{\rm d}\tau{\rm
  d}\rho+\frac{B^2-C^2}{B} {\rm d}\rho^2+ \rho^2({\rm d}\theta^2+\rm 
sin\theta^2 {\rm d}\phi^2). \label{le2} 
\end{equation} 
Such a break down of the new coordinate system cannot, however, occur
in a neighbourhood of the centre where, in virtue of (\ref{c}) and
(\ref{eEtr}), $B+C= 1$ holds there identically.

Finally, we would like to emphasise that to get the
simple form of the evolution equation (\ref{c}) -- (\ref{eEtr}) for
our basic variables $B, C, \Phi$ and $\Pi$ we could also start with
the metric (\ref{le2}) and apply the substitutions 
\begin{equation}
\frac {\partial}{\partial \rho} \psi=\Phi + \frac CB \Pi,\ \ \ \ 
\frac {\partial }{\partial \tau}\psi= \frac 1B \Pi. \label{sub2}
\end{equation}
These later relations can be derived by making use of (\ref{sub1}) and
(\ref{s2}) -- (\ref{r2}). Notice also that in this setup there is no
need to solve any equation for $A$ since the time evolution of all the
basic variables is governed by (\ref{c}) -- (\ref{eEtr}).  

\section{Final remarks}
\setcounter{equation}{0}

The favourable aspects of the use of the Kodama vector field  in the
investigated  evolutionary problem gets apparent if one thinks of the
straightforward process yielded the above simple setup. In
fact, it is very unlikely that whenever one is given the investigated
gravity scalar field system, without making use of the Kodama vector
field, one would ever chose the gauge in which the metric possesses
the form (\ref{le2}), moreover, one would also apply the substitution
(\ref{sub2}).

It is also important to emphasise that the above described process had 
immediately provided a system of first order symmetric hyperbolic 
evolution equations for the smallest possible number of variables. The
significance of this point gets to be transparent whenever, in case of
other gauge choices, one tries to use the standard techniques of
reduction to get a system of first order PDEs from second order
ones. It might be surprising but, in general, such a  simple minded
process does not 
automatically yield a system of strongly hyperbolic evolution
equations. Recall that without possessing this property even the
well-posedness of the associated evolutionary problem is not
guaranteed.

It is also worth mentioning  that the main conclusions of the present
paper generalise straightforwardly to the case of self-interacting
matter fields. In the appendix the evolution equations relevant for
case of a generic self-interacting scalar field, along with the
related modifications of the argument applied above for the massless
scalar field can be find. The time evolution of more complicated
coupled gravity-matter systems, including various non-linearly coupled
gauge fields like Yang-Mills and Higgs fields, will be studied
elsewhere. 

There is a definite disadvantage of the applied canonical gauge
representation, which is, in fact, shared by any of the gauge
representations using the area radial coordinate. The disadvantage is
that the trapped region cannot be represented in an associated
coordinate domain since it necessarily breaks down at the marginally
trapped surfaces. In particular, in case of the applied gauge choice
the marginally trapped surfaces can only appear in the limit
$B=(1-2m/r)^{-1}\rightarrow \infty$.

{\bf Acknowledgements} I wish to thank to Hideo Kodama  for stimulating
discussions. Thanks are also in due to the Yukawa Institute for
Theoretical Physics where parts of the investigations were carried
out during the Post-YKIS2005 scientific program for its kind
hospitality. This research was also supported in parts by OTKA grant
T034337. The author is a Bolyai János Research Fellow of the 
Hungarian Academy of Sciences. 

\section*{Appendix}

\renewcommand{\theequation}{A.\arabic{equation}}
\setcounter{equation}{0}

Consider now the particular case of a self-interacting scalar field
$\Psi$, i.e. we shall assume that the dynamics of $\Psi$ can be
derived from a Lagrangian of the form ${\mathcal L}=\nabla^e \Psi
\nabla_e \Psi + V(\Psi)$, where the potential $V$ is an arbitrary but
sufficiently regular expression depending only on the field variable
$\Psi$. For instance, $V(\Psi)$ could be chosen to be the widely used
quadratic potential, $V(\Psi)=\lambda \left(\Psi^2-\Psi_0^2
\right)^2$, with $\lambda, \Psi_0 \geq 0$, although as it will be seen
below the following argument does not require such a limitation in the
functional form of $V(\Psi)$.

The field equation and the energy-momentum tensor relevant for the
above specified generic system read as 
\begin{equation}
\nabla^e \nabla_e \Psi-\frac12\frac{\partial V}{\partial \Psi}=0,\label{we2}
\end{equation}
and 
\begin{equation}
T_{ab}=\nabla_a \Psi \nabla_b \Psi -\frac12 g_{ab}\left( \nabla^e \Psi
  \nabla_e \Psi +V(\Psi)\right). \label{Tab2} 
\end{equation}
A direct calculation also justifies that the associated changes in
equations (\ref{11st}), (\ref{22nd}), (\ref{A}) and (\ref{const}) can
be given as  
\begin{equation}
{\frac {\partial }{\partial \tau }}\,\Phi ={\displaystyle 
\frac { \,B\,({\frac {\partial }{\partial \rho }}\Pi )\, - 
C({\frac {\partial }{\partial \rho }}\Phi )}{ B^{2} - C^{2}}} +
{\displaystyle \frac {B(B -1)\Pi  -  C (B+1)\Phi } {\rho 
    ( B^{2} - C^{2})}}+ {\displaystyle \frac {4\pi\rho B 
    (C\Phi -B\Pi) V +\frac12 BC({\frac
  {\partial V}{\partial \Psi }})} {B^{2} - C^{2}}}\,,\label{111st}  
\end{equation}
\begin{equation}
{\frac {\partial }{\partial \tau }}\,\Pi ={\displaystyle \frac {
 B({\frac {\partial }{\partial \rho }}\Phi ) - C({\frac {
\partial }{\partial \rho }}\Pi )}{ B^{2} - C^{2}}}
+ {\displaystyle \frac {B(B +1)\Phi  -C (B-1)\Pi } {\rho( B^{2} -
    C^{2})}}
- {\displaystyle \frac {4\pi\rho B
    (B\Phi-C\Pi) V +\frac12 B^2({\frac
  {\partial V}{\partial \Psi }})} {B^{2} - C^{2}}}\,
, \label{222nd}
\end{equation}
\begin{equation}
\,{\frac {\partial }{\partial \rho }}\,{\rm ln}A - C\,{\frac {
\partial }{\partial \tau }}\,{\rm ln}A  - \frac{(\,B  -1)}{\rho}  - 4\,\pi
\,\rho\,\left(\,\Phi ^{2}+ \,\Pi ^{2}-B\, V\right)=0\,, \label{AA}
\end{equation}
\begin{equation}
\frac {\partial }{\partial \rho }\,B=4\,\pi\,\rho  \,\left(B\,(\Phi ^{2} + 
\Pi ^{2}\,+\, B\, V) + 2\,C\,\Pi\,\,\Phi \right) - {\displaystyle \frac {B\,(B
 - 1)}{\rho}}. \label{cconst}
\end{equation}

Notice first that the metric variable $A$ does not entered to either of the
formerly $A$-independent equations. In fact, the most important change
is that not only the derivatives $\frac {\partial }{\partial
\tau}\Psi$ and $\frac {\partial }{\partial \rho }\Psi$, given in terms
of $\Pi$ and $\Phi$, enter to the above field equations but also the
field $\Psi$ itself via the self-interaction potential $V(\Psi)$. This
necessitates, in addition to (\ref{c}), (\ref{111st}), (\ref{222nd})
and (\ref{eEtr}), the use of the evolution equation
\begin{equation}
{\frac {\partial }{\partial \tau }}\,\Psi=\frac \Pi B\label{Phitau}
\end{equation}  
for $\Psi$. This equation can be derived from the relations defining
$\Pi$ and that of (\ref{r1}), moreover, (\ref{Phitau}) is also independent of
the metric variable $A$. More importantly, (\ref{Phitau}), (\ref{111st}), 
(\ref{222nd}), (\ref{eEtr}) and (\ref{c}) can be seen to possess the
form of a first order symmetric hyperbolic system 
\begin{equation}
\frac{\partial}{\partial \tau}u=A_\rho\left(\frac{\partial}{\partial
  \rho}u\right)+S(u) ,\label{hyp}
\end{equation}
where the vector variable $u$ and the symmetric coefficient matrix
$A_\rho$ are given as  
\begin{equation}
u=\left(
\begin{array}{c}
{\Psi } \\
{ \Pi} \\
\Phi \\
B \\
C \\
\end{array}
\right),\ \ {\rm and} \ \ 
A_\rho= \left(
\begin{array}{ccccc}
0 & 0 & 0 & 0 & 0  \\
0 & -\frac{C}{B^2-C^2} & \frac{B}{B^2-C^2} & 0 & 0  \\
0 & \frac{B}{B^2-C^2} & -\frac{C}{B^2-C^2} & 0 & 0  \\
0 & 0 & 0 & 0 & 0 \\
0 & 0 & 0 & 0 & 0 \\
\end{array}
\right), 
\end{equation}
moreover, the vector expression $S(u)$ stores all the suitably
arranged source 
terms of equations (\ref{Phitau}), (\ref{111st}), (\ref{222nd}),
(\ref{eEtr}) and (\ref{c}).  For such a system the initial value
problem is known to be well-posed. Moreover, again we have that this
system -- consisting of (\ref{Phitau}), 
(\ref{111st}), (\ref{222nd}), (\ref{eEtr}) and (\ref{c}) --, along
with (\ref{AA}), comprises a complete 
system, i.e. (\ref{cconst}) and $E_{\theta\theta}=0$ can be derived
from them, whenever (\ref{Phitau}), (\ref{111st}), (\ref{222nd}),
(\ref{eEtr}), (\ref{c}) and (\ref{AA}) hold everywhere, moreover,
(\ref{cconst}) is satisfied on the initial data surface. Since, as it
was before, (\ref{Phitau}), (\ref{111st}), (\ref{222nd}),
(\ref{eEtr}) and (\ref{c}) separate from (\ref{AA}) the later
can be solved for the metric variable $A$ independently upon the
functions $\Psi, \Pi, \Phi, B $ and $C$ have been determined.

\vfill\eject

\end{document}